\documentclass[aps,prl,twocolumn,showpacs,preprintnumbers,amsmath,amssymb,superscriptaddress,longbibliography]{revtex4-2}
\usepackage{hyperref}
\hypersetup{colorlinks=true, citecolor=blue, urlcolor=blue, linkcolor=blue,urlcolor=cyan}
\usepackage{color}
\usepackage{graphicx}
\usepackage{dcolumn}
\usepackage{braket}

\def\ep{{\epsilon}}

\def\frac#1#2{{#1\over #2}}

\def\s{\sqrt}

\def\be{\begin{equation}}
\def\ee{\end{equation}}
\def\ba{\begin{eqnarray}}
\def\ea{\end{eqnarray}}
\def\de{\partial}

\def\ti{\tilde}
\def\ap{\alpha}

\def\la{\langle}
\def\lb{\rangle}
\def\ep{\epsilon}

\newcommand{\eg}{{\it e.g.,}\ }
\newcommand{\cf}{{\it cf.}\ }
\newcommand{\ie}{{\it i.e.,}\ }
\newcommand{\viz}{{\it viz.,}\ }

\newcommand{\mt}[1]{\textrm{\tiny #1}}
\renewcommand{\(}{\left(}
\renewcommand{\)}{\right)}

\newcommand{\Scon}{S^{\mathrm{con}}}
\newcommand{\Sdis}{S^{\mathrm{dis}}}
\newcommand{\GN}{G_{\mt{N}}}

\begin{document}

\title{On the Page curve under final state projection}
\preprint{YITP-21-158; IPMU21-0086}

\author{Ibrahim Akal}\email{ibrakal@yukawa.kyoto-u.ac.jp}
\affiliation{\it Center for Gravitational Physics, Yukawa Institute for Theoretical Physics, Kyoto University, Kitashirakawa Oiwakecho, Sakyo-ku, Kyoto 606-8502, Japan}

\author{Taishi Kawamoto}\email{taishi.kawamoto@yukawa.kyoto-u.ac.jp}
\affiliation{\it Center for Gravitational Physics, Yukawa Institute for Theoretical Physics, Kyoto University, Kitashirakawa Oiwakecho, Sakyo-ku, Kyoto 606-8502, Japan}

\author{Shan-Ming Ruan}\email{ruan.shanming@yukawa.kyoto-u.ac.jp}
\affiliation{\it Center for Gravitational Physics, Yukawa Institute for Theoretical Physics, Kyoto University, Kitashirakawa Oiwakecho, Sakyo-ku, Kyoto 606-8502, Japan}

\author{Tadashi Takayanagi}\email{takayana@yukawa.kyoto-u.ac.jp}
\affiliation{\it Center for Gravitational Physics, Yukawa Institute for Theoretical Physics, Kyoto University, Kitashirakawa Oiwakecho, Sakyo-ku, Kyoto 606-8502, Japan}
\affiliation{\it Inamori Research Institute for Science, 620 Suiginya-cho, Shimogyo-ku, Kyoto 600-8411, Japan}
\affiliation{\it Kavli Institute for the Physics and Mathematics of the Universe,University of Tokyo, Kashiwa, Chiba 277-8582, Japan}

\author{Zixia Wei}\email{zixia.wei@yukawa.kyoto-u.ac.jp}
\affiliation{\it Center for Gravitational Physics, Yukawa Institute for Theoretical Physics, Kyoto University, Kitashirakawa Oiwakecho, Sakyo-ku, Kyoto 606-8502, Japan}

\begin{abstract}
The black hole singularity plays a crucial role in formulating Hawking's information paradox. The global spacetime analysis may be reconciled with unitarity by imposing a final state boundary condition on the spacelike singularity. Motivated by the final state proposal, we explore the effect of final state projection in two dimensional conformal field theories. We calculate the time evolution under postselection by employing the real part of pseudo-entropy to estimate the amount of quantum entanglement averaged over histories between the initial and final states. We find that this quantity possesses a Page curve-like behavior.
\end{abstract}
\maketitle

\begin{figure}[ht!]
	\centering
	\includegraphics[width=3.2in]{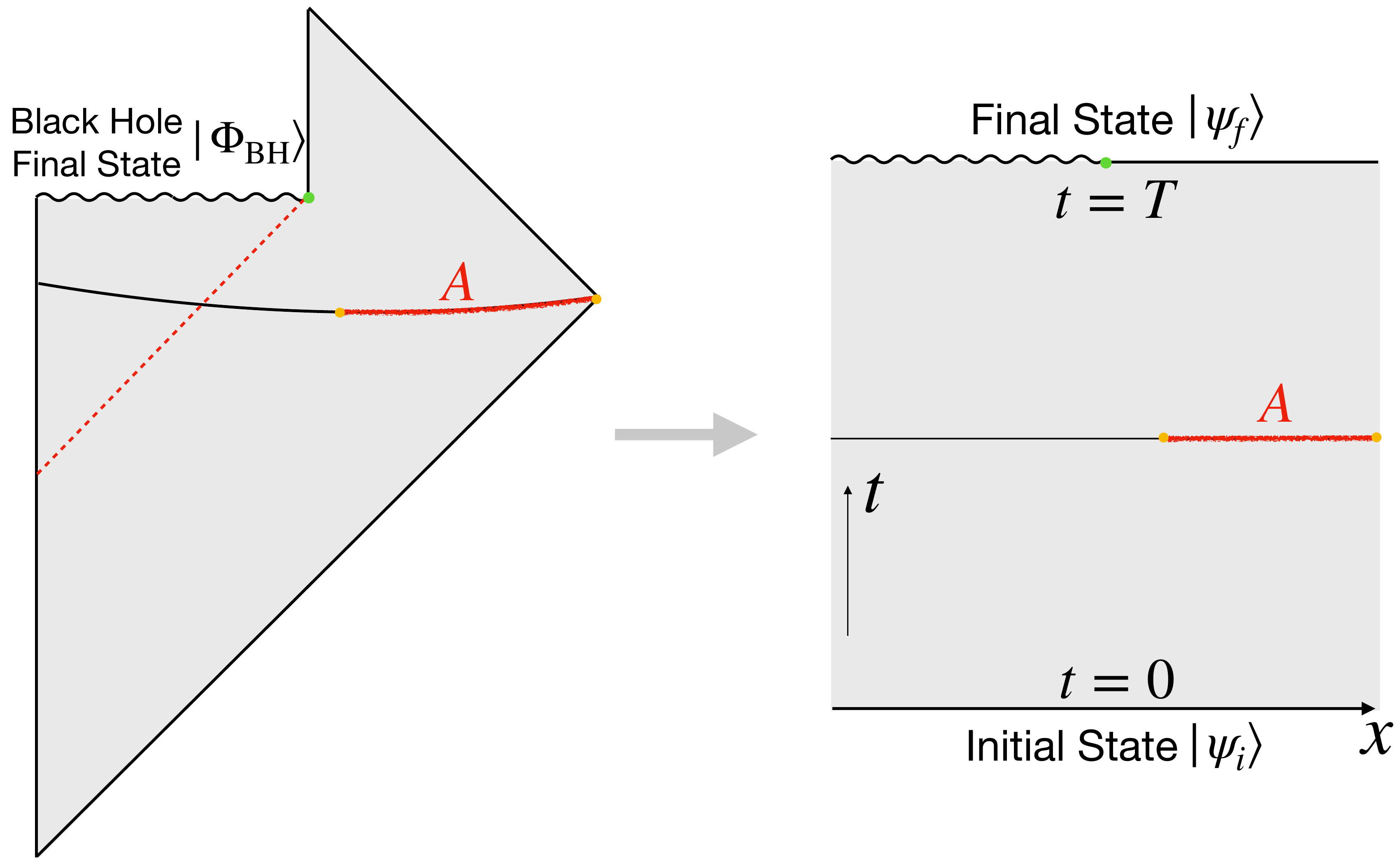}
	\caption{Final state projection in an evaporating black hole (left) and its field theory simplification (right).}
	\label{fig:setupbh}
\end{figure}

\noindent \emph{1.~Introduction.---}The process of postselection is useful in studies of dynamical properties of quantum many-body systems and quantum field theories (QFTs). The nonunitary dynamics of projective measurements, for instance, provides a new tool for controlling many-body systems, giving rise to measurement-induced phase transitions \cite{Skinner:2018tjl,Li:2018mcv}. Postselection also plays a key role in the black hole final state proposal \cite{Horowitz:2003he} that suggests a possible resolution to the black hole information puzzle. Namely, even though the evaporation process due to Hawking radiation \cite{Hawking:1974rv,Hawking:1975vcx} seems to change the initial pure state forming the black hole by gravitational collapse into a mixed state \cite{Hawking:1976ra}, the final state is still a pure state due to the state projection imposed on the spacelike singularity, \cf the left panel of Fig.~\ref{fig:setupbh}. However, it has been pointed out that the black hole final state has to be of a particular type to preserve information \cite{Gottesman:2003up}.

On the other hand, it has earlier been argued by Page that black hole unitarity should be reflected in the entanglement spectrum of Hawking radiation \cite{Page:1993df,Page:1993wv}. The basic idea is that the Bekenstein-Hawking entropy \cite{Bekenstein:1973ur,Hawking:1975vcx} gives the leading term for the logarithm of the number of black hole microstates. Accordingly, when the number of states of the radiation becomes larger than that of the remaining microstates, the entanglement entropy between the two should be upper bounded by the Bekenstein-Hawking entropy of the black hole. The curve that describes the time evolution of the entanglement entropy is known as the Page curve.

It is essential to mention that studies of postselection in QFTs have been quite limited until so far. One reason for this may be the lack of universal and calculable quantities that can characterize the evolution of quantum states in the corresponding processes. For example, quantum quenches are often studied as a typical class of time-dependent systems, and entanglement entropy is important in probing how the systems thermalize \cite{Calabrese:2005in}.

However, in the presence of postselection, the use of entanglement entropy is limited---nevertheless, see \cite{Rajabpour:2015uqa,rajabpour2015fate,Rajabpour:2015xkj,Numasawa:2016emc}---since it only depends on a single quantum state, namely either on the initial state or the final state. Instead, it is desirable to consider a universal quantity reflecting histories from the initial state to the final state under postselection.

Recently, one such candidate, called pseudo-entropy, has been introduced in \cite{Nakata:2020luh} and defined as follows. Let us consider two pure states $|\psi_1\lb$ and $|\psi_2\lb$, and decompose the total system into the subsystems $A$ and $B$ such that the entire Hilbert space becomes factorized into ${\cal H}_{\rm tot}={\cal H}_A\otimes {\cal H}_B$. Taking the reduced transition matrix as 
\begin{equation}
\tau^{1|2}_A=\mbox{Tr}_B\left[\frac{|\psi_1\lb \la \psi_2|}{\la \psi_2|\psi_1\lb}\right]\,.
\end{equation}
the pseudo-entropy is given by
\begin{equation}\label{PEdef}
S^{1|2}_A=-\mbox{Tr}\,[\tau^{1|2}_A\log\tau^{1|2}_A]\,.
\end{equation}
This quantity arises as a straightforward extension of holographic entanglement entropy \cite{Ryu:2006bv,Hubeny:2007xt} in the case of Euclidean time-dependent backgrounds \cite{Nakata:2020luh}. Pseudo-entropy is generically complex-valued as the transition matrix is not always Hermitian. However, we expect that the real part of pseudo-entropy, \ie $\mathrm{Re}[S^{1|2}_A]$, can be understood as the number of Bell pairs averaged over histories evolving from $|\psi_1\lb$ to $|\psi_2\lb$. For more explanation, see appendix A. Thus, this quantity is equally sensible in figuring out the underlying correlation structure as in the case of entanglement entropy. Furthermore, the mentioned quantity provides a nice quantum order parameter to distinguish different quantum phases \cite{Mollabashi:2020yie, Mollabashi:2021xsd}. Refer to \cite{Camilo:2021dtt,Nishioka:2021cxe,Goto:2021kln,Miyaji:2021lcq} for further recent progress related to pseudo-entropy.

The purpose of this article is to uncover the time evolution of pseudo-entropy in conformal field theories (CFTs) under postselection, mainly motivated by the black hole final state proposal, as depicted in the left panel of Fig.~\ref{fig:setupbh}.
More specifically, we consider an initial state $|\psi_i\lb$ at time $t=0$, which evolves under the Hamiltonian $H$ until $t=T$ when we perform the postselection to the final state $\ket{\psi_f}$. At time $t$, the initial state evolves into 
$|\psi_1\lb=e^{-iHt}\ket{\psi_i}$, while the final state is reverted to $|\psi_2\lb=e^{i(T-t)H}\ket{\psi_f}$. In this setup, sketched in the right panel of Fig.~\ref{fig:setupbh}, we can define the pseudo-entropy as in Eq.~\eqref{PEdef}. This is the quantity we will study in this paper. Note that
if we apply the replica method by assuming a cut along the subsystem $A$, the familiar quantity given by $-\frac{\de}{\de n}\log \frac{Z_n}{(Z_1)^n}$, where $Z_n$ is the partition function on the $n$-sheeted geometry, actually coincides with the pseudo-entropy, but not entanglement entropy for either $|\psi_1\lb$ or $|\psi_2\lb$. This quantity reduces to the entanglement entropy only when $|\psi_1\lb=|\psi_2\lb$. 
Here, we often assume that the CFT under our consideration has a classical gravity dual in order to obtain analytical results. Such a CFT, \ie a so-called holographic CFT, is strongly coupled and has a large central charge $c$ \cite{Heemskerk:2009pn,Hartman:2014oaa}. For this class of CFTs, we can evaluate correlation functions by employing the large $c$ factorization such that they can be computed by Wick contractions.


\vspace{4 pt}
\noindent \emph{2.~Homogeneous postselection and gravity dual.---}Let us start with the simplest model where we perform a postselection homogeneously. A tractable class of postselection can be defined by using the boundary state $\ket{B}$ (or Cardy state \cite{Cardy:1989ir}) and considering the following two pure states in a given CFT
\begin{equation}\label{globalstate}
\begin{split}
     \ket{\psi_1}=e^{-iHt}e^{-\delta H}\ket{B}\,,\, \ket{\psi_2}=e^{iH(T-t)}e^{-\delta H}\ket{B}\,, 
\end{split}
\end{equation}
where the parameter $\delta$ denotes a UV regularization of the boundary state. The inner product of the initial state and final state defines a Euclidean path-integral on the strip with the width $L=2\delta+iT$. We can describe this strip as $0\leq \text{Im}\,w\leq L$ by taking $(w,\bar{w})$ as a complex coordinate. Performing the conformal transformation as follows
\begin{equation}\label{eq:conformalmapsimple}
z=e^{\frac{\pi w}{L}}\,,
\end{equation}
one can map this strip into an upper half-plane defined by Im$\,z>0$. 

We set the two end points $w_1, w_2$ of the interval $A$ as $(w_i,\bar{w}_i)=(x_i+i(\delta+it), x_i-i(\delta+it))$ with $i=1,2$. 
The $n$-sheeted partition function $Z_n$ can be computed by inserting the twist operator $\sigma_n$ at the two end points of $A$ as what is usually done in the standard computation of entanglement entropy in field theories \cite{Calabrese:2004eu}. In holographic CFTs, this two-point function has two different saddle point contributions, namely, (i) the Wick contraction of two twist operators and (ii) the Wick contraction of each twist operator with their mirror images. Since the candidates of pseudo-entropy computed from (i) and (ii) are dual to the length of the connected and disconnected geodesic, we refer to them as the connected and disconnected contribution, denoted by $\Scon_A$ and $\Sdis_A$, respectively.

More generally, if we consider a conformal map from the original $w$ coordinate to the upper half-plane in $z$ coordinate via a holomorphic map $z=f(w)$, the two distinct contributions to pseudo-entropy are given by  
\begin{align}
S^{\mathrm{con}}_{A}&=\frac{c}{6}\log\frac{|f(w_1)-f(w_2)|^2}{\ep^2 |f'(w_1)||f'(w_2)|} \,,\label{eq:sconsimp} \\
S^{\mathrm{dis}}_A&=\frac{c}{6}\log\frac{|f(w_1)-\bar{f}(\bar{w}_1)||f(w_2)-\bar{f}(\bar{w}_2)|}{\ep^2 |f'(w_1)||f'(w_2)|}
+2S_{\mathrm{bdy}}\,,\label{eq:sdissimp}
\end{align}
where $S_{\mathrm{bdy}}$ is referred to as the boundary entropy \cite{Affleck:1991tk} and $\epsilon$ is a UV cutoff corresponding to lattice spacing. This field-theoretic result perfectly agrees with the holographic entanglement entropy in AdS/BCFT (anti-de Sitter/boundary CFT) \cite{Takayanagi:2011zk,
Fujita:2011fp,Karch:2000gx}.

Using the conformal map \eqref{eq:conformalmapsimple} for the present model under a homogeneous postselection, we can find
\begin{equation}\label{eq:PEsimple}
\begin{split}
\Scon_A&=\frac{c}{3}\log \left[\frac{2T}{\pi\ep}\sin\left(\frac{\pi (x_2-x_1)}{2T}\right)\right], \\
\Sdis_A &=\frac{c}{3}\log \left[\frac{2T}{\pi\ep}\sin\left(\frac{\pi t }{T}\right)\right]+ i\,\frac{\pi c}{6}+2S_{\rm bdy}\,,
\end{split}
\end{equation}
by taking the limit $\delta\to 0$.
The value of pseudo-entropy is given by 
the one which has a smaller real part, \viz
\begin{equation}\label{minPE}
\mbox{Re}\,[S^{1|2}_A]=\mbox{min}\left[\mbox{Re}\,[\Scon_A],\mbox{Re}\,[\Sdis_A]\right]\,.
\end{equation}
Note that $\mbox{Re}\,[\Sdis_A]$ vanishes at $t=\ep/2$, which we regard as the initial time with a regularization. Then it increases logarithmically, reaching the maximum at the middle time $t=T/2$. It again decreases and vanishes at the final time $t=T-\ep/2$ as the boundary state does not have any real space entanglement \cite{Miyaji:2014mca}. On the other hand, $\mbox{Re}\,[\Scon_A]$ is time-independent and is identical to the entanglement entropy for the vacuum state. Taking into account the minimization in the definition \eqref{minPE}, it is clear that the disconnected contribution $\mbox{Re}\,[\Sdis_A]$ dominates at early and late times. Depending on the value of $S_{\rm bdy}$, there is a chance that $\mbox{Re}\,[\Scon_A]$ dominates for a finite period $t_\ast<t<T-t_\ast$.

\begin{figure}[ht!]
	\centering
	\includegraphics[height=2.1in]{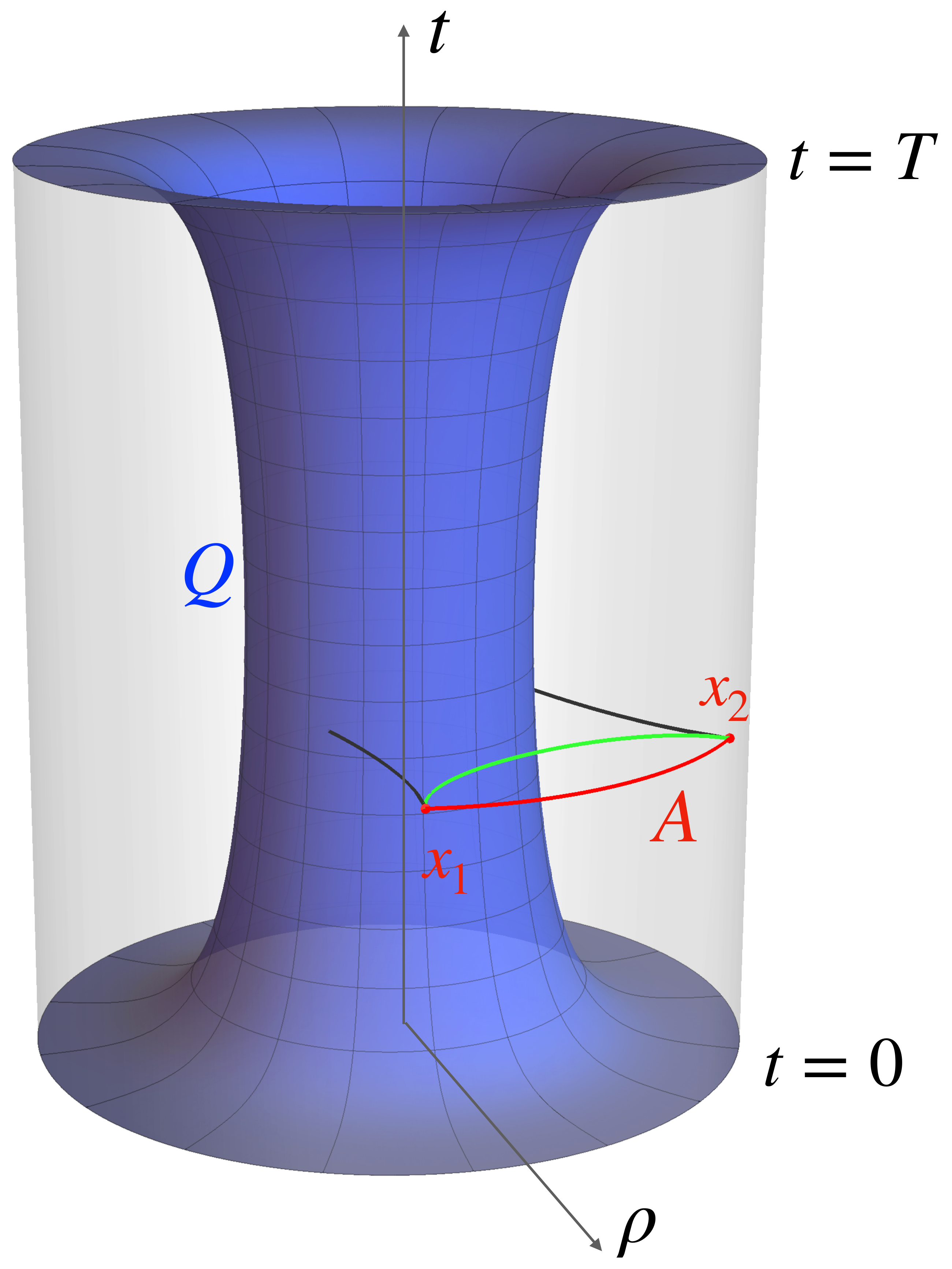}
	\caption{The gravity dual of the Lorentzian BCFT. The blue surface describes the EOW brane defined in Eq.~\eqref{eq:adsthmetq}. For a subsystem A (red curve) located on the asymptotic boundary, the black and green curves denote the corresponding disconnected geodesics and the connected geodesic, respectively.}
	\label{fig:adsthqd}
\end{figure}

The holographic analysis \cite{Akal:2020wfl} based on AdS/BCFT suggests that a spacelike boundary in a Lorentzian BCFT has a complex-valued boundary entropy, namely
\begin{equation}\label{eq:bdyent}
S_{\rm bdy}=\frac{c}{6}\log\s{\frac{|{\cal T}|-1}{|{\cal T}|+1}}-i\,\frac{\pi c}{12}\,, 
\end{equation} 
where ${\cal T}$ is the tension of the end-of-the-world (EOW) brane dual to the boundary of the BCFT.
This takes values in the range ${\cal T}<-1$. The imaginary part of \eqref{eq:bdyent} exactly cancels that of the pseudo-entropy $\Sdis_A$. 

For our setup associated with \eqref{globalstate}, we can further construct its gravity dual as follows.
Considering a global AdS$_3$, \ie
\begin{equation}\label{eq:adsthmet}
ds^2=-\frac{T^2}{\pi^2}\cosh^2\rho \,dt^2+d\rho^2+\frac{T^2}{\pi^2}\sinh^2\rho \,dx^2\,,
\end{equation}
we introduce an EOW brane $Q$ defined by 
\begin{equation}\label{eq:adsthmetq}
\cosh\rho\sin\frac{\pi t}{T}=\cosh\eta_0\,,
\end{equation} 
that describes two-dimensional de Sitter spacetime.
Here $\eta_0$ is related to the brane tension as
${\cal T}=-\coth\eta_0$. Finally, the gravity dual of the present CFT setup is given by the region surrounded by the
AdS asymptotic boundary $\rho\to\infty$ and the EOW brane $Q$, as illustrated in Fig.~\ref{fig:adsthqd}.
Similar to holographic entanglement entropy \cite{Ryu:2006bv,Hubeny:2007xt}, the holographic pseudo-entropy  \cite{Nakata:2020luh} is given by the geodesic length $L_A$ in the corresponding Lorentzian gravity dual. In our case, due to the presence of $Q$, the geodesic connecting two endpoints of the interval $A$ can end on $Q$, as shown in Fig.~\ref{fig:adsthqd}. Thus, we can have contributions from disconnected geodesics, denoted by $L^{\rm dis}_A$, in addition to those resulting from a connected geodesic $L^{\rm con}_A$. 
Accordingly, the real part of holographic pseudo-entropy is taken to be the minimum as before, \ie
\ba
\mbox{Re}\,[S^{1|2}_A]=\mbox{min}\left[\mbox{Re}\left[\frac{L^{\rm con}_A}{4\GN}\right],
\mbox{Re}\left[\frac{L^{\rm dis}_A}{4\GN}\right]\right]\,.
\ea

Obviously, the connected contribution is the same as the holographic entanglement entropy in global AdS$_3$ and thus coincides with Eq.~\eqref{eq:sconsimp}. The connected contribution arises only when the connected geodesic does not touch the EOW brane $Q$, which leads to the following condition
\begin{equation} \label{eq:condtoc}
\sin\frac{\pi t}{T}>\cosh\eta_0\cdot \sin\frac{\pi (x_2-x_1)}{2T}\,.
\end{equation}
Moreover, a straightforward computation for the sum of two geodesic lengths reproduces the disconnected contribution as shown in Eqs.~\eqref{eq:sdissimp} and \eqref{eq:bdyent}. In this way, we can derive the CFT results from the gravity dual. More details are presented in appendix B.

\begin{figure}[ht!]
  \centering
   \includegraphics[width=8cm]{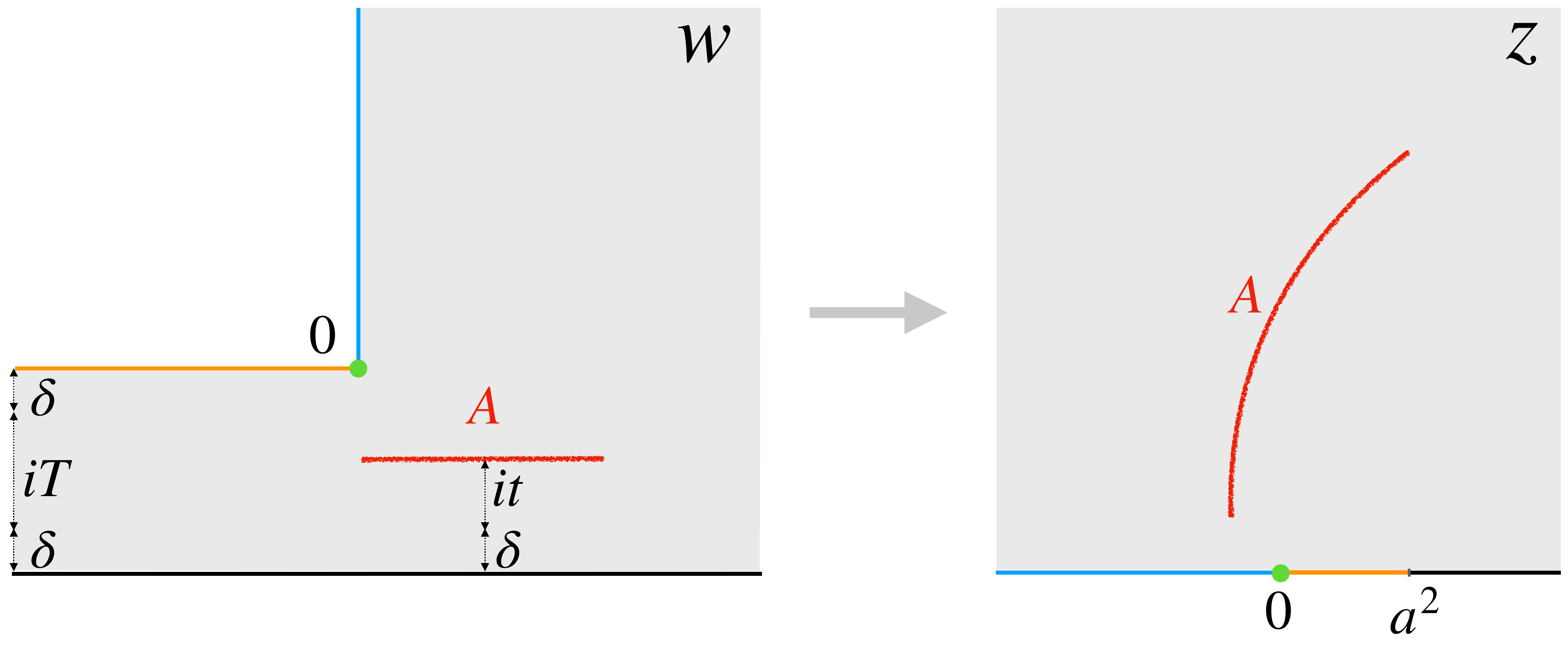}
  \caption{A sketch of Euclidean path-integral on the $w$-sheet for inhomogenous postselection and its conformal transformation to the upper half-plane in terms of the $z$ coordinate. For $x<0$, the path-integral is terminated at $\text{Im}\,w=0$ describing the projection to $\ket{B}_{x<0}$, while 
  for $x>0$ it extends to $\text{Im}\,w=\infty$, corresponding to the projection to the vacuum state $\ket{0}_{x>0}$ on the right half.}\label{fig:qqqEEP}
\end{figure}

\vspace{4 pt}
\noindent \emph{3.~Inhomogeneous postselection and pseudo-entropy.---}Next, we focus on an example of inhomogeneous postselection models motivated by the black hole final state projection. Namely, at $t=0$
we project the left part $x<0$ and right part $x>0$ to a boundary state $\ket{B}$ and the CFT vacuum $\ket{0}$, respectively. This is realized by the (Euclidean) path-integral on the $w$-sheet as depicted in the left of Fig.~\ref{fig:qqqEEP}.
We regard the projection on $x<0$ as the black hole final state \footnote{One may wonder whether the boundary state $\ket{B}$ imposed on $x<0$ can mimic the black hole final state since the latter is expected to be very random \cite{Horowitz:2003he}. However, it is worth emphasizing that, as commented in \cite{Gottesman:2003up}, the state which is expected to be random should be a time-reverted state of the black hole final state, but not itself. Therefore, there is no apparent contradiction that a boundary state can mimic the black hole final state.}.
Although the boundary state is based on a local boundary condition and does not have any real space entanglement \cite{Miyaji:2014mca}, we expect that the time evolution $e^{i(T-t)H}$ may lead the state to a random chaotic one. We also qualitatively mimic the creation of entangled pairs due to Hawking radiation during the time evolution of the initial boundary state as similar to quantum quenches \cite{Calabrese:2005in}.

Taking the initial state as the quantum quench state $e^{-\delta H}\ket{B}$, and fixing the final state as $e^{-\delta H}\ket{B}_{x<0}\otimes \ket{0}_{x>0}$, we can calculate the pseudo-entropy by using the general formulae shown in Eqs.~\eqref{eq:sconsimp} and \eqref{eq:sdissimp}. We choose the subsystem $A$ as an interval whose two endpoints are denoted by $(w_1,\bar{w}_1)$ and $(w_2,\bar{w}_2)$. More explicitly, one can find ($i=1,2$)
\begin{equation}\label{eq:wwbar}
    w_i=x_i+i \(-\delta - i (T-t) \)\,,  \bar{w}_i=x_i - i \(-\delta - i (T-t) \)\,.
\end{equation}
We can also map the $w$-sheet to an upper half-plane ($z$-plane) via the following conformal transformation
\begin{equation}
\begin{split}
w &=\s{z}-\frac{a}{2}\log\left[\frac{1+\frac{\s{z}}{a}}{\frac{\s{z}}{a}-1}\right]-\frac{a}{2}\pi i\,,\\
\bar{w}&=\s{\bar{z}}
-\frac{a}{2}\log\left[\frac{1+\frac{\s{\bar{z}}}{a}}{\frac{\s{\bar{z}}}{a}-1}\right]+\frac{a}{2}\pi i\,, \\
\end{split}
\end{equation}
by choosing $a=\frac{2}{\pi}\(2\delta+ i T\)$.

For simplicity, we first examine the simple cases with $|x|\gg T$ to obtain some analytical results. To wit, one can get
$z \simeq  (x-t)^2,\  \bar{z}\simeq (x+t)^2\,,$
and 
$z\simeq a^2\!+\!4a^2 e^{\frac{2}{a}(x-t)}\,, \bar{z}\simeq a^2\!+\!4a^2 e^{\frac{2}{a}(x+t)}\,,$
by taking the limit $x\gg T$ and $x\ll -T$, respectively. It is then straightforward to evaluate the pseudo-entropy in the following three cases: 
(a) $x_1,x_2\gg T$, (b) $x_2\gg T, x_1\ll -T$, and (c) $x_1,x_2\ll -T$ by setting $\delta=0$ and noting that $t<T\ll |x_{1,2}|$.

In the case (a), $x_1,x_2\gg T$, we obtain
\begin{equation}
\begin{split}
\mbox{Re}\,[\Scon_A]&\simeq \frac{c}{6}
\log\left[ \frac{(x_1-x_2)^2 (x_1+x_2)^2}{4 x_1 x_2\ep^2}\right]\,,\\
\mbox{Re}\,[\Sdis_A]&\simeq \frac{c}{3}
\log\left[\frac{2t}{\ep}\right]+2\ti{S}_{\rm bdy}\,,
\end{split}
\end{equation}
where $\ti{S}_{\rm bdy}$ is the real part of the boundary entropy $S_{\rm bdy}$ for space-like surface.
At $t=0$, the disconnected one is favored and we have $\mbox{Re}\,[\Sdis_A]=0$.
As time evolves, it grows logarithmically as 
$\mbox{Re}\,[\Sdis_A]\simeq \frac{c}{3}\log \frac{2t}{\ep}$ and 
eventually $\mbox{Re}\,[\Scon_A]$ becomes dominant.
This transition behavior is schematically described by 
\begin{equation}\label{transph}
\mbox{Re}\,[S_A]\simeq \min \left[\frac{c}{3}\log \frac{|x_1-x_2|}{\ep},\frac{c}{3}\log \frac{2t}{\ep}+2\ti{S}_{\rm bdy}\right]\,. 
\end{equation}
We can intuitively understand this result as follows. The state $\ket{\psi_i}=e^{-itH}e^{-\delta H}\ket{B}$ has quantum entanglement for the length scale $l<2t$ due to causal propagation. Thus, if $|x_1-x_2|<2t$, both the initial state and final states have the corresponding quantum entanglement.
However, if $|x_1-x_2|>2t$, the initial state does not have the entanglement at the length scale $|x_1-x_2|$. These arguments qualitatively explain the behavior of Eq.~\eqref{transph}. It is still intriguing that the growth is given in terms of the logarithmic one as opposed to the linear growth of entanglement entropy in global quenches. 

In the case (b), $x_2\gg T$ and $x_1\ll -T$, we obtain
\begin{equation}
\begin{split}
\mbox{Re}\,[\Scon_A]&\simeq \frac{c}{6}
\log\left[\frac{\pi^2 x_2^3}{32T\ep^2}\right]\,,\\
\mbox{Re}\,[\Sdis_A]&\simeq \frac{c}{6}
\log\left[\frac{4Tt\sin\left(\frac{\pi t}{T}\right)}
{\pi \ep^2}\right]+2\ti{S}_{\rm bdy}\,.
\end{split}
\end{equation}
Since we assume $x_2\gg T>t$, the disconnected one is favored at any time.
It starts with $\mbox{Re}\,[\Sdis_A]=0$ as in the previous case and grows logarithmically as $\mbox{Re}\,[\Sdis_A]\simeq \frac{c}{3}\log \frac{2t}{\ep}$ initially. Then it reaches a maximum and starts decreasing.
At the final time $t\simeq T$, it is reduced to $\mbox{Re}\,[\Sdis_A]\simeq \frac{c}{6}\log \frac{2T}{\ep}$. 
This result can be explained by noting that 
the entanglement in the part $x_1<x<0$ of region $A$ becomes trivial at the final time due to postselection. 

In the case (c), $x_1,x_2\ll -T$, we simply reproduce our previous results \eqref{eq:sconsimp} and \eqref{eq:sdissimp}. This can be easily understood if we note that for $x\ll -T$, the space looks like a strip having width $T$, identical to the homogeneous postselection model \eqref{globalstate}.

\begin{figure}[ht!]
	\centering
	\includegraphics[width=4.2cm]{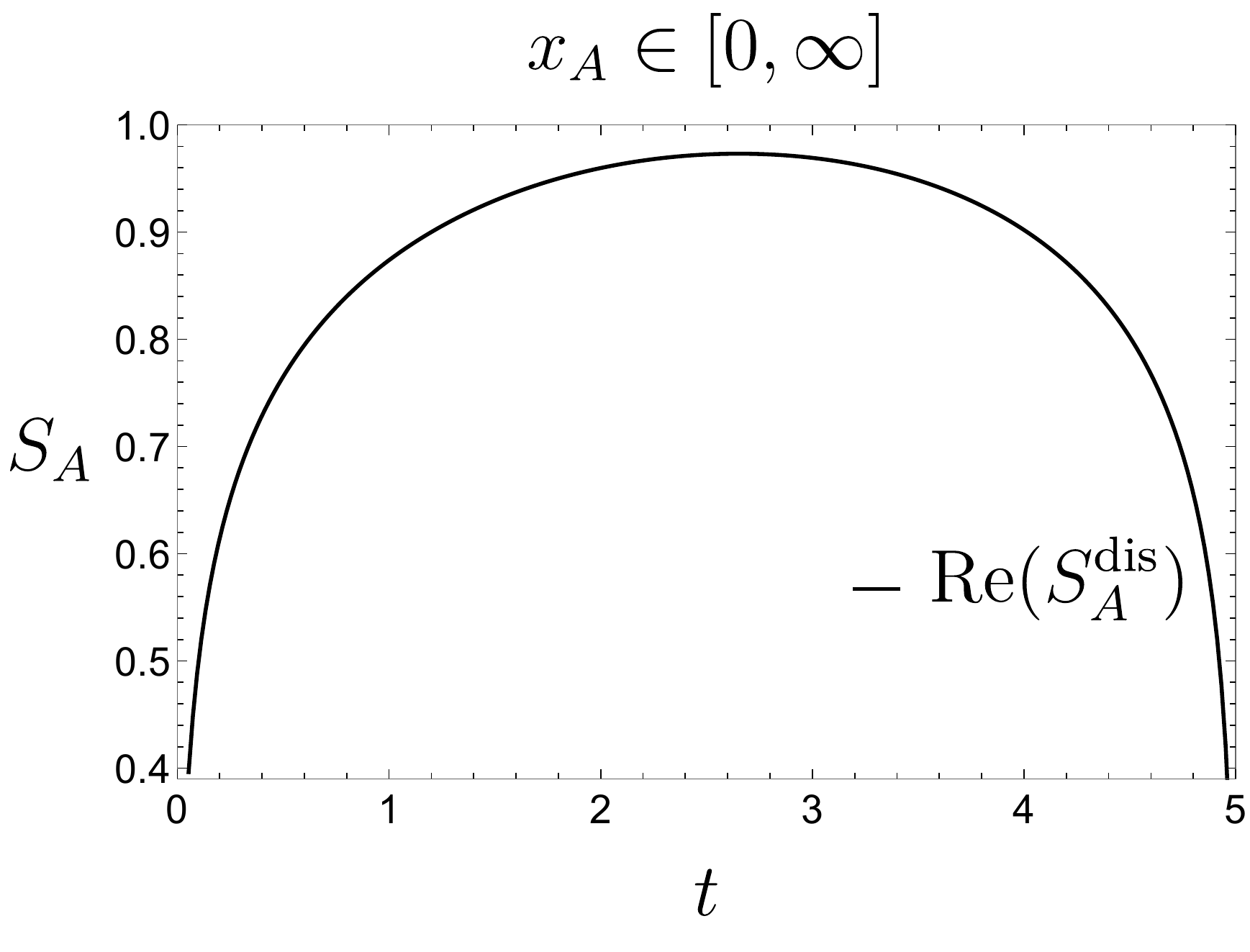}
	\includegraphics[width=4.2cm]{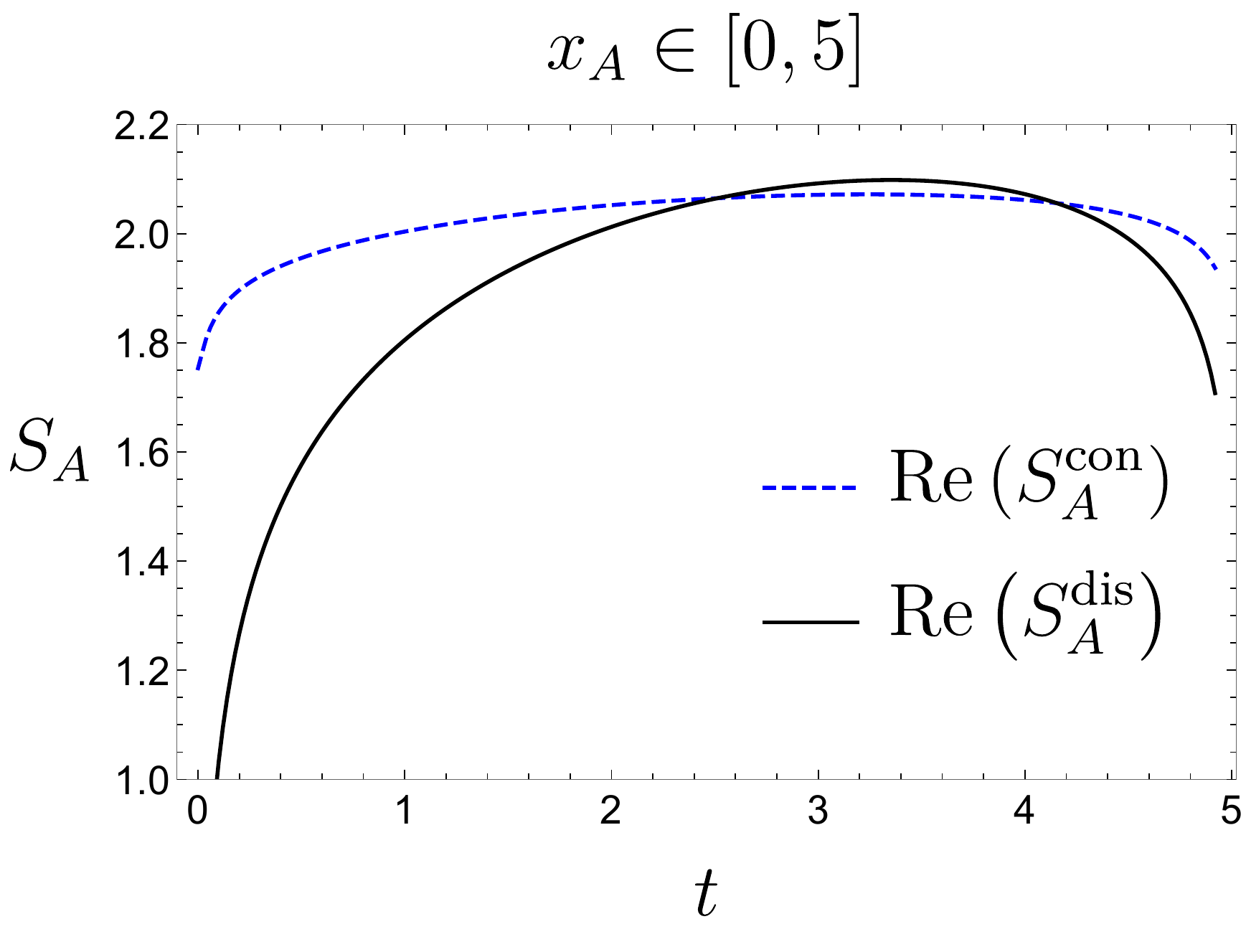}
	\caption{The plot for the real part of pseudo-entropy Re$[S_A]$ for subsystems $A=[0,\infty]$ (left) and $A=[0,5]$ (right) as a function of time $t$. We choose $T=5$, $\delta=\ep=0.01$ and $\ti{S}_{\rm bdy}=0$.}
	\label{fig:pagepejce}
\end{figure}

Next, we shall choose the subsystem to be $A=[0,y]$ with $y>0$ to model the radiation subsystem for the evaporating black hole, where the black hole final state is imposed on $x<0$ at $t=T$.  We numerically plot the real part of pseudo-entropy in Fig.~\ref{fig:pagepejce}. In particular, when we choose $y=\infty$, where only the disconnected geodesic is available, it looks like a Page curve, \ie starting from $\mbox{Re}\,[S_A]=0$ and ending up with $\mbox{Re}\,[S_A]=0$. Indeed, since we have $w\simeq -\frac{z^{3/2}}{3a^2}$ when $z\simeq 0$, we can estimate the value at $t\simeq T$ as follows
\ba
\mbox{Re}\,[\Sdis_A]\sim \frac{c}{12}\log \frac{|w|^2}{\ep^2}\simeq  \frac{c}{6}\log \frac{(T-t)^2+\delta^2}{\ep^2}\,.
\ea
At $t=T$, this is of the same order as at the initial time $t=0$. Therefore, we can conclude that the pseudo-entropy vanishes at $t=T$ by choosing $\delta=O(\ep)$. More generally, this causes the phenomenon that the real part of pseudo-entropy is highly reduced at $|x|=T-t$. We may interpret that this is due to the negative energy flux emitted at the end point of the projection region, $x=0$, at $t=T$.


\begin{figure}[h!]
	\centering
	\includegraphics[width=3.2in]{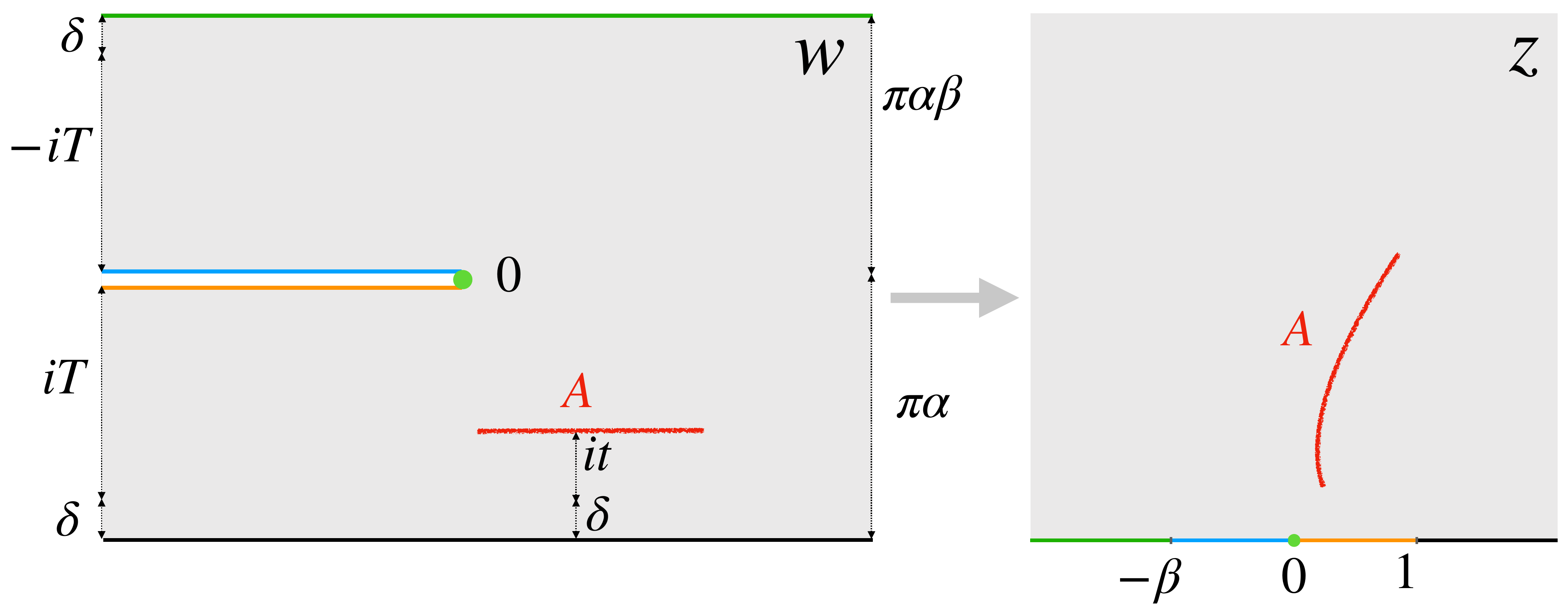}
	\caption{A sketch of the path-integral description of partial postselection (left) and its conformal map to the upper half-plane (right).}
	\label{fig:PE}
\end{figure}

\vspace{4 pt}
\noindent \emph{4.~Partial postselection and pseudo-entropy.---}Although the inhomogeneous postselection model studied before mimics a black hole spacetime with a spacelike singularity, there is a crucial difference with the black hole final state proposal \cite{Horowitz:2003he}. In the former, the postselection is only imposed on the singularity, while no operation is performed to the compliment part. 

To model this feature, we shall consider partial postselection in a two-dimensional CFT, where we make a projection only for the left half $x<0$. We again choose a global quench state $e^{-\delta H}|B\lb$ at $t=0$ as the initial state and consider its time evolution until the partial postselection is imposed at $t=T$. 
We take the postselected state as a boundary state $|B\lb_{x<0}$, while keeping the right half $x>0$ free. Therefore, right after the postselection at $t=T$, the quantum state associated with the whole system turns out to be 
\begin{align}
    &\,|B\lb_{x<0} \otimes \left[\la B|_{x<0}~ \left(e^{-iTH} e^{-\delta H} |B\lb\right) \right]_{x>0} \nonumber\\
    = & \left(|B\lb \la B|_{x<0}\otimes I_{x>0}\right)  \cdot e^{-iTH} e^{-\delta H} |B\lb\,,
\end{align}
where $I_{x<0}$ is the identity matrix for the left half.

\begin{figure}[ht!]
  \centering
    \includegraphics[width=1.6in]{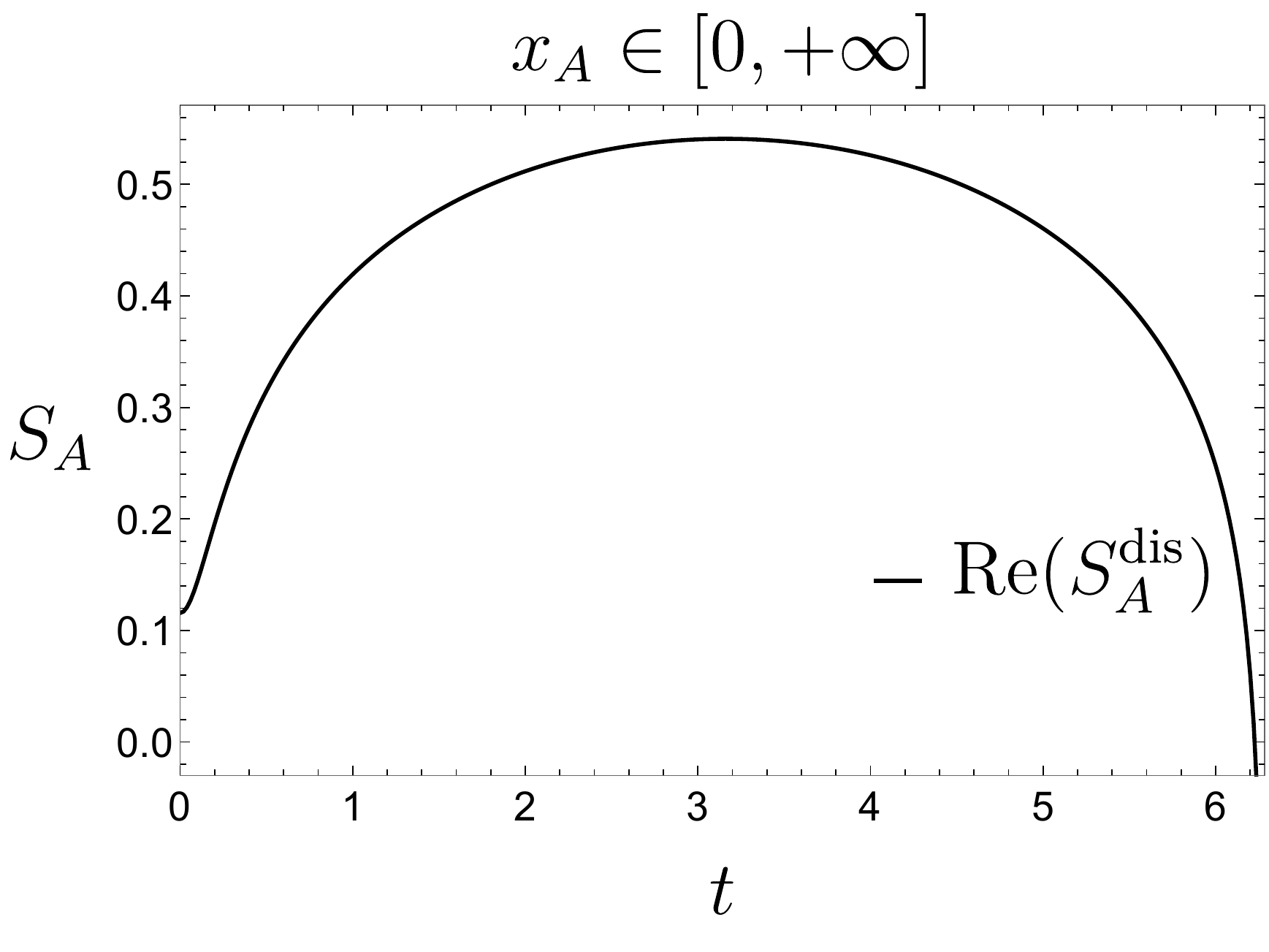}
     \includegraphics[width=1.6in]{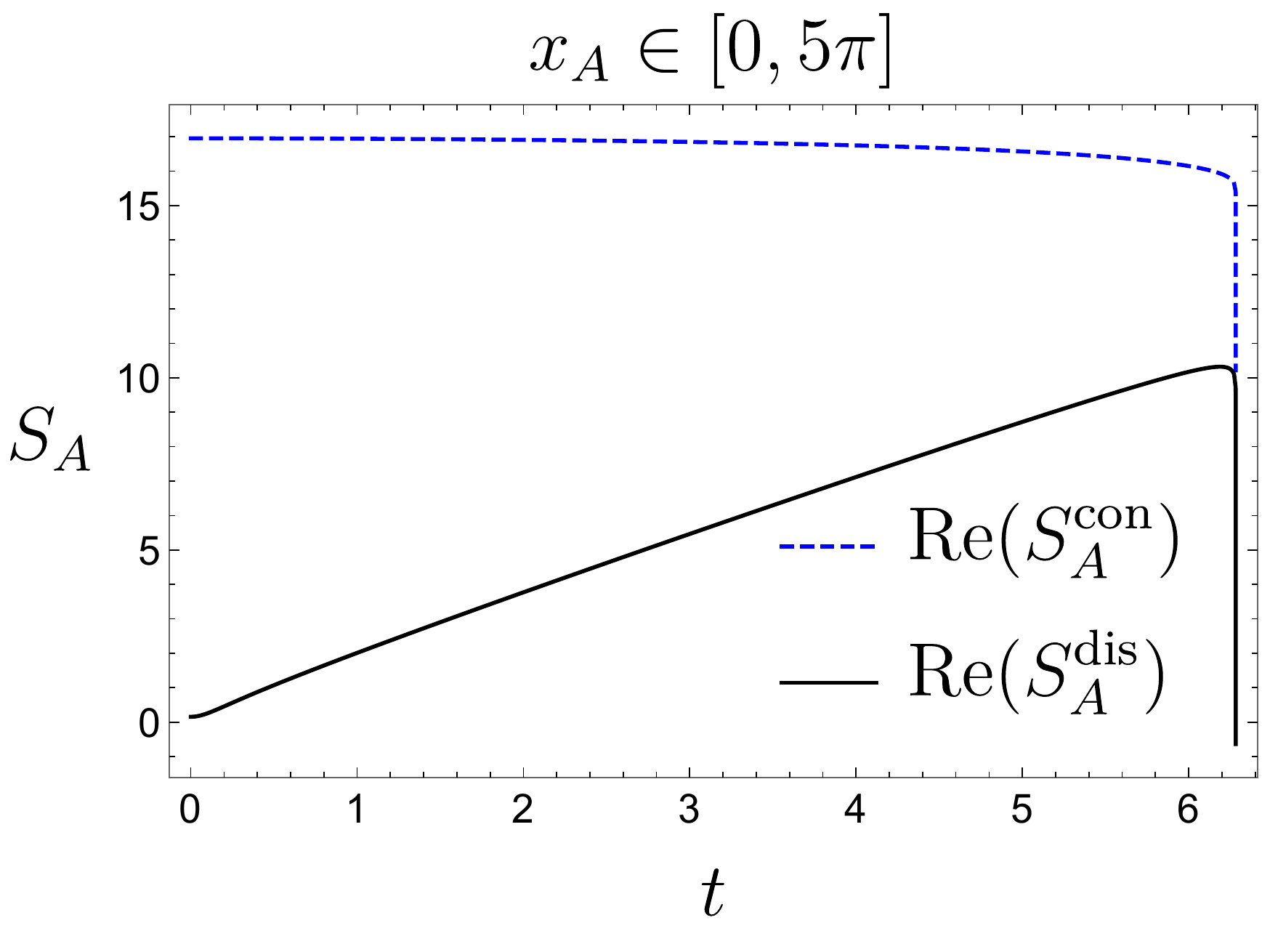}
  \caption{The plot for the real part of pseudo-entropy Re$[S_A]$ 
  for subsystems $A=[0,\infty]$ (left) and $A=[0,5\pi]$ (right) as a function of time $t$. We choose the parameters $T=2\pi$, $\delta=\varepsilon=0.05 \pi$.}
\label{fig:PEzpage}
\end{figure}

Next, we focus on the pseudo-entropy at immediate time $0<t<T$. In this case, the initial state and the final state are given by 
\begin{equation}
\begin{split}
|\psi_1\lb &= e^{-itH} e^{-\delta H} |B\lb\,,\\
    |\psi_2\lb &= e^{-i(t-T)H} \left(|B\lb \la B|_{x<0}\otimes I_{x>0}\right)\cdot e^{-iTH} e^{-\delta H} \ket{B}\,.
\end{split}
\end{equation}
Taking subsystem $A$ to be an interval, the pseudo-entropy (\ref{PEdef}) in this setup can be computed via the Euclidean path-integral shown in the left panel of Fig.~\ref{fig:PE}. Using the conformal map
\begin{equation}
w(z)=f^{-1}(z)=\ap \log(1-z)+\ap\beta\log\left(\frac{z+\beta}{\beta}\right)\,,
\end{equation}
we can map the strip geometry with a slit to the upper half-plane as shown in the right of Fig.~\ref{fig:PE}. The parameters $\alpha$ and $\beta$ are fixed by $\pi\alpha=\delta+iT$ and $\pi\alpha\beta=\delta-iT$, respectively. For the subsystem $A=[0,y]$, with $y>0$, we plot the real part of pseudo-entropy in Fig.~\ref{fig:PEzpage}. This again shows a Page curve-like behavior.

\vspace{4 pt}
\noindent \emph{5.~Discussion.---}We have studied the time evolution of pseudo-entropy under postselection, whose real part provides an estimation of the amount of quantum entanglement, \ie the number of Bell pairs averaged over histories between the initial state and the postselected final state. In a two-dimensional CFT setup, where postselection applies to the left region, $x<0$, the mentioned entanglement measure associated with the region $x>0$ possesses a Page curve-like behavior. It grows by starting from zero, reaches its maximum, decreases, and eventually vanishes at the time of the partial postselection.
Our setups model black hole evaporation according to the final state projection scenario. One may wonder how these models are related to the island picture \cite{Penington:2019npb,Almheiri:2019psf,Almheiri:2019hni} resulting in the Page curve. 
Notably, in our analysis, the unitarity-caused deviation from monotonical increase arises due to the past evolution of the postselected final state. This qualitatively looks similar to the island scenario suggesting a modification of the Hilbert space structure inside the black hole.
However, state projection seems to affect correlations already in the initial stages of evaporation. There is no sudden appearance of a disjoint entanglement region on a Cauchy slice residing in the interior right after the Page time.

Considerations that seem to indicate a tension between the final state proposal and the presence of a smooth horizon have been discussed in \cite{Lloyd:2013bza,Bousso:2013uka}, where a resolution to the puzzle raised in \cite{Bousso:2013uka} has been proposed in \cite{AlmheiriTalk}. Nevertheless, starting from the arguments in \cite{Almheiri:2012rt}, we shall emphasize that indications of a drama for infalling observers are merely an artifact of the factorizable Hilbert space within the improper semiclassical treatment.
On the other hand, the notion of a classical singularity may already have to be given up when the black hole is young, following the arguments by Page \cite{Akal:2021csd}. This motivates considering sequential projective measurements scanning through a smeared interior patch. 
It might be possible that associated correlations responsible for the Page curve make it to the exterior near-horizon region, namely in a nontrivially protected form \cite{Akal:2020ujg}. A correct treatment of the singularity might thus have imprints not only behind the horizon. We want to return to some of these aspects in the near future. Another interesting direction is studying state projection in moving mirrors \cite{Akal:fsp-hmm} by employing their holographic formulation \cite{Akal:2020twv,Akal:2021foz}.

\vspace{8 pt}
\begin{acknowledgments}
\emph{Acknowledgments.---}We are grateful to Tokiro Numasawa and Kotaro Tamaoka for useful discussions and especially to Ahmed Almheiri for valuable comments on a draft of this paper. We would also like to thank the YITP workshop: ”KIAS-YITP 2021: String Theory and Quantum Gravity ” (YITP-W-21-18), hosted by YITP, Kyoto U, for stimulating discussions and comments from participants, where this work was presented by T.\,K. This work is supported by MEXT-JSPS Grant-in-Aid for Transformative Research Areas (A) "Extreme Universe", No.\,21H05187. T.\,T. is also supported by JSPS Grant-in-Aid for Scientific Research (A) No.\,21H04469.
S-M.\,R. and T.\,T. are supported by the Simons Foundation through the ``It from Qubit'' collaboration. T.\,T. is also supported by Inamori Research Institute for Science and World Premier International Research Center Initiative (WPI Initiative) from MEXT. 
Z.\,W. is supported by Grant-in-Aid for JSPS Fellows
No. 20J23116. 

\end{acknowledgments}

\bibliography{PEpage.bib}

\pagebreak
\titlepage
\setcounter{page}{1}

\section{Appendix A: Pseudo-entropy and postselection}\label{app:A}

In the following, we present an argument suggesting that the real part of pseudo entropy can measure quantum entanglement when we take an average over the evolution from the initial state to the final state under postselection.

Let us consider the decoherence function \cite{Gell-Mann:1989dby} for the projective measurement $\sum_k \Pi_k=1$ defined by
\ba
D_{k,l}=\frac{1}{\mbox{Tr}[\rho_1\rho_2]}\mbox{Tr}[\rho_2 \Pi_k  \rho_1\Pi_l]\,,
\ea
where $\rho_1$ and $\rho_2$ denote the initial and final density matrix, respectively. This quantity measures the inference of probability of $k$-event and $l$-event. When the decoherence function is diagonal, the histories completely decohere. 
If we set $\rho_2=I$, the above quantity is reduced to the ordinary probability 
distribution
\begin{equation}
D_{k,l}=\mbox{Tr}\,[\Pi_k\rho_i\Pi_l]\,.
\end{equation}
When both the initial and final state are pure, we have 
\begin{equation}
    D_{k,l}=\frac{\la \psi_f|\Pi_k|\psi_i\lb\la\psi_i|\Pi_l|\psi_f\lb}{|\la\psi_f|\psi_i\lb|^2}\,.
\end{equation}

As in \cite{Nakata:2020luh}, let us consider two-qubit states $|\psi_1\lb$ and $|\psi_2\lb$ of the following form
\begin{equation}\label{eq:bell}
\begin{split}
|\psi_1\lb&=c_1|00\lb_{AB}+s_1|11\lb_{AB}\,,\\
|\psi_2\lb&=c_2|00\lb_{AB}+s_2|11\lb_{AB}\,,
\end{split}
\end{equation}
where $c_{1,2}$ and $s_{1,2}$ can take any complex values with the constraints $|c_1|^2+|s_1|^2=1$ and  $|c_2|^2+|s_2|^2=1$ imposed.
To have a better counting of Bell pairs, 
we take the asymptotic limit $M\to\infty$,  by considering $M$ copies of the original states: $|\psi_i\lb=(|\psi_1\lb)^{\otimes M}$ and 
$|\psi_f\lb=(|\psi_2\lb)^{\otimes M}$. The total Hilbert space now consists of $2M$ qubits, \ie $M$ copies of the $A$ spin and $M$ copies of the $B$ spin. We call the former $\ti{A}$ and the latter $\ti{B}$. 
We choose $\Pi_k$ to be the projection which acts only on the $M$ spins in $\ti{A}$ such that the states with $k$ up spins (\ie $|1\lb$) and $M-k$ down spins (\ie $|0\lb$) for $M$-qubit states are selected. $\Pi_k$ acts on $\ti{B}$ as an identity. After the projection by $\Pi_k$, we obtain a state with maximal entanglement between $\ti{A}$ and $\ti{B}$. Due to this projection there remain ${}_MC_k\equiv \frac{M!}{k!(M-k)!}$ states. This is the same procedure as in \cite{Nakata:2020luh}, where an operational interpretation of pseudo-entropy was presented for this special class of states.

We naturally define the averaged value $\bar{N}$ of Bell pairs when we fix both the initial and final state in the asymptotic limit as follows
\begin{equation}\label{www}
    \bar{N}=\lim_{M\to \infty}\frac{N_{\rm max}}{M}\,,
\end{equation}
where 
\ba
N_{\rm max}=\sum_{k,l}D_{k,l}
\left(\frac{\log ({}_MC_k) +\log ({}_M C_l)}{2}\right)\,. \label{wwww}
\ea
Here, $N_{max}$ estimates the maximal number of Bell pairs, which can be distilled by local operation and classical communication (LOCC) in the asymptotic limit $M\to\infty$ when we take the average over histories 
from the initial state to the final state.
We can explicitly write 
\ba
D_{k,l}=p_k p^*_l\,,
\ea
with
\ba
p_k=\frac{\la \psi_f|\Pi_k|\psi_i\lb}{\la \psi_f|\psi_i\lb}\,.
\ea
Note that $p_k$ take complex values in general since we allow $c_{1,2}$ and 
$s_{1,2}$ to take complex values.
Due to the identity $\sum_k p_k=1$, we can rewrite $N_{\rm max}$ as 
\begin{equation}
N_{\rm max}=\mbox{Re}\left[\sum_{k} p_k \log ({}_MC_k)\right]\,.
\end{equation}
As shown in \cite{Nakata:2020luh}, the quantity $\sum_{k} p_k \log ({}_MC_k)$ coincides with 
the pseudo-entropy (\ref{PEdef}) in the limit $M\to \infty$. Therefore, it implies that the real part of pseudo-entropy can be interpreted as the number of distillable Bell pairs averaged over the history between the initial state and final state. In this argument, we have assumed the specific class of initial and final states given by (\ref{eq:bell}). They share the same basis of spins, \ie $|00\lb_{AB}$ and $|11\lb_{AB}$, and we can easily fix the form of the projection $\Pi_k$. To extend this analysis to general states, we need to pick up an appropriate projection $\Pi_k$ to extract Bell pairs, which is not obvious for generic choices of $\ket{\psi_1}$ and $\ket{\psi_2}$. We will leave this general argument to future work.

\section{Appendix B: Analysis of end-of-the-world branes in $\text{AdS}_3$ }\label{adsthddd}

We focus on the end-of-the-world (EOW) brane in global AdS$_3$ \eqref{eq:adsthmet} defined by \eqref{eq:adsthmetq}. This surface is described by its world-sheet coordinates $(\tau,x)$ introduced as follows
\begin{equation}
    \begin{split}
        X_0&=\cosh\rho\cos\frac{\pi t}{T}=\sinh\tau \sinh\eta_0\,,\\
X_1&=\sinh\rho\sin\frac{\pi x}{T}=\cosh\tau\sinh\eta_0\sin\frac{\pi x}{T}\,,\\
X_2&=\sinh\rho\cos\frac{\pi x}{T}=\cosh\tau\sinh\eta_0\cos\frac{\pi x}{T}\,,\\
X_3&=\cosh\rho\sin\frac{\pi t}{T}=\cosh\eta_0\,,\\
    \end{split}
\end{equation}
where the original AdS$_3$ defined by $X_0^2+X_3^2=X_1^2+X_2^2+1$ in the spacetime with line element
$ds^2=-(dX_0)^2-(dX_3)^2+(dX_1)^2+(dX_2)^2$.
The induced metric of the brane $Q$ is derived as
\ba
ds^2|_Q=\sinh^2\eta_0\left(-d\tau^2+\frac{\pi^2}{T^2}\cosh^2\tau dx^2\right),
\ea
which is nothing but a two-dimensional de Sitter space.
 
To evaluate the geodesic length, it is useful to rewrite the global AdS$_3$ in terms of Poincare 
coordinates, \ie 
\begin{equation}
\begin{split}
X_0&=\frac{1+z_{\mt{P}}^2+x_{\mt{P}}^2-t_{\mt{P}}^2}{2z_{\mt{P}}}\,,\\
X_3&=\frac{t_{\mt{P}}}{z_{\mt{P}}}\,,\\
X_1&=\frac{x_{\mt{P}}}{z_{\mt{P}}}\,,\\
X_2&=\frac{1-z_{\rm{P}}^2-x_{\mt{P}}^2+t_{\mt{P}}^2}{2z_{\mt{P}}}\,,\\
\end{split}
\end{equation}
with the metric $ds^2=\frac{dz_{\mt{P}}^2-dt_{\mt{P}}^2+dx_{\mt{P}}^2}{z_{\mt{P}}^2}$.
The surface $Q$ mapped to the plane is thus located at 
\ba
\frac{t_{\mt{P}}}{z_{\mt{P}}}=\cosh\eta_0\,.
\ea
It is also useful to note that we have $x_{\mt{P}}=\frac{\sin\frac{\pi x}{T}}{\cos\frac{\pi t}{T}+\cos\frac{\pi x}{T}}$ and $t_{\mt{P}}=\frac{\sin\frac{\pi t}{T}}{\cos\frac{\pi t}{T}+\cos\frac{\pi x}{T}}$ at the AdS boundary . 

In Poincare coordinates, the geodesic length between a boundary point at $t_{\mt{P}}$ and the surface $Q$ is given by $\log\left[\frac{2t_P}{\ep}e^{\eta_0}\right]$ (refer to \eg \cite{Akal:2020wfl}). As a result, we can reproduce Eq.~\eqref{eq:sdissimp}. The condition that the connected geodesic connecting two boundary points $(t_{\mt{P}},x^{(1)}_{\mt{P}})$ and $(t_{\mt{P}},x^{(2)}_{\mt{P}})$ does not touch the surface $Q$ is given by
\begin{equation}
|x^{(2)}_{\mt{P}}-x^{(1)}_{\mt{P}}|<2z^*_{\mt{P}}\,,
\end{equation}
where $z^*_{\mt{P}}$ denotes the value of $z_{\mt{P}}$ at the time $t_{\mt{P}}$ on the surface $Q$, \ie $z^*_{\mt{P}}=\frac{t_{\mt{P}}}{\cosh\eta_0}$. This explains the condition given in Eq.~\eqref{eq:condtoc}.

\end{document}